\documentclass[final,5p,times,twocolumn]{elsarticle}

\usepackage{graphicx}

\usepackage{amssymb}
\usepackage{amsmath}
\usepackage{color}

\journal{Physics Letters B}
\newcommand{\bea}{\begin{eqnarray}}
\newcommand{\eea}{\end{eqnarray}}
\newcommand{\bi}{\begin{itemize}}
\newcommand{\ei}{\end{itemize}}
\newcommand{\benu}{\begin{enumerate}}
\newcommand{\eenu}{\end{enumerate}}
\newcommand{\nn}{\nonumber}
\newfont{\bg}{cmr10 scaled\magstep4}

\begin{document}

\begin{frontmatter}
\title{Higgs Production in Two-Photon Process and Transition Form Factor}

\author[KEK]{Norihisa Watanabe}
\ead{norihisa@post.kek.jp}

\author[KEK]{Yoshimasa Kurihara}
\ead{kurihara@post.kek.jp}

\author[yokohama]{Ken Sasaki}
\ead{sasaki@ynu.ac.jp}

\author[kyoto]{Tsuneo Uematsu}
\ead{uematsu@scphys.kyoto-u.ac.jp}

\address[KEK]{High Energy Accelerator Research Organization (KEK)\\
 Tsukuba, Ibaraki 305-0801, Japan}

\address[yokohama]{ Dept. of Physics, Faculty of Engineering\\
 Yokohama National University, Yokohama 240-8501, Japan}

\address[kyoto]{Institute for Liberal Arts and Sciences,
Kyoto University, Kyoto 606-8501, Japan\\
and Maskwa Institute, Kyoto Sangyo University, Kyoto 603-8555, Japan}

\begin{abstract}
The Higgs production in the two-photon fusion process 
is investigated 
where one of the photons is off-shell while the other one is on-shell.
This process is realized in either electron-positron collision or electron-photon collision 
  where the scattered electron or positron 
is detected (single tagging) and described by the transition form 
factor.
We calculate the contributions to the transition form factor of the Higgs
boson coming from top-quark loops 
and W-boson loops. We then study the $Q^2$ dependence of each
contribution to the total transition form factor
 and also of  the differential cross 
section for the Higgs production. 
\end{abstract}


\end{frontmatter}


\section{Introduction \label{introduction}}
There has been much interest in the diphoton decay of the Higgs
boson discovered at LHC experiments \cite{HiggsLHC}, since its coupling to
the photon is connected with the question whether it is really
a Standard Model (SM) Higgs boson
or the one beyond SM, such as in
the minimal supersymmetric standard model (MSSM) or in composite models.
It would be intriguing to
investigate the properties of the SM Higgs boson
through the 
production process in the two-photon fusion:\ $2\gamma\rightarrow H$, 
which might be realized at ILC~\cite{ILC} and is just the opposite reaction of the diphoton decay mode of the 
Higgs boson:\ $H\rightarrow 2\gamma$. The Higgs diphoton decay goes through
charged fermion loops and 
W-boson loops as discussed in Ref.\cite{Ellis,Ioffe,Shifman,Rizzo,Gavela, Marciano,Hunter}
and the references therein.
\begin{figure}[hbt]
\begin{center}
\includegraphics[scale=0.25]{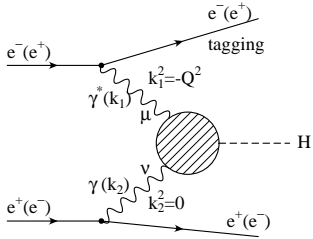}
\caption{\label{eecollision} $e^+~e^-$ two-photon fusion 
process for the Higgs production.
} 
\end{center}
\end{figure}

\begin{figure}[hbt]
\begin{center}
\includegraphics[scale=0.28]{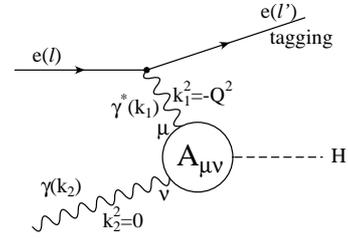}
\caption{\label{egammacollision} $e~\gamma$ two-photon fusion 
process for the Higgs production.
} 
\end{center}
\end{figure}

Here we particularly interested in the  virtual and real two-photon processes 
(i) the electron-positron collision in Fig.\ref{eecollision}, where one of 
the scattered electron (or positron) is detected (single tagging), and (ii) 
the electron-photon collision $e\gamma\rightarrow eH$ shown in 
Fig.\ref{egammacollision} where we observe the scattered electron. From 
these processes we 
can measure the so-called \lq\lq transition form factor\rq\rq of 
the Higgs boson as a function of the virtual-photon mass squared~\footnote{
The $\gamma^*\gamma \rightarrow \pi^0$ 
transition form factor was first  investigated in QCD~\cite{BrodskyLepage}.
The recent experimental data were given in Refs.\cite{BABAR, Belle} . }.
\vspace{-0.3cm}

In this paper we investigate the SM Higgs boson production in the  virtual and real two-photon fusion $\gamma^*\gamma \rightarrow H$
shown in Figs.\ref{eecollision} and \ref{egammacollision} and calculate the transition form factor of the 
Higgs boson at one-loop level.
First we examine the tensor structure of the transition amplitude
for $\gamma^*\gamma\rightarrow H$  which respects gauge invariance.
We then evaluate the contributions to the amplitude
from charged fermion loops and $W$-boson loops.
\section{Higgs production and transition amplitude}
The transition amplitude for $\gamma^*\gamma\rightarrow H$ extracted from the process in  Fig \ref{egammacollision} is given by
\bea
M\equiv\langle H|T|\gamma^*(k_1)\gamma(k_2)\rangle
=\epsilon^\mu(k_1)\epsilon^\nu(k_2)\ A_{\mu\nu}(k_1,k_2),
\eea
where $\epsilon^\mu(k_1)$ ($\epsilon^\nu(k_2)$) is the polarization
vector of the incident virtual (real) photon, $k_1^2=-Q^2<0$ and  $k_2^2=0$. 
Due to the electromagnetic gauge invariance,
the tensor $A_{\mu\nu}$  can be decomposed as
\bea
&&\hspace{-0.5cm}A_{\mu\nu}(k_1,k_2)
=\left(g_{\mu\nu}(k_1\cdot k_2)
-k_{2\mu}k_{1\nu}\right)S_1(m^2,Q^2,m_H^2)\nn\\
&&\hspace{1cm}+\left(k_{1\mu}k_{2\nu}-\frac{k_1^2}{k_1\cdot k_2}k_{2\mu}k_{2\nu}\right)
S_2(m^2,Q^2,m_H^2),
\eea
where $m_H$ is the Higgs boson mass satisfying
$(k_1+k_2)^2=m_H^2$ and the  intermediate particle masses in the loop are 
collectively denoted by $m$.
Since $k_2^\nu \epsilon_\nu(k_2)=0$, the transition 
amplitude reads
\bea
&&\hspace{-0.5cm}M
=\left[g^{\mu\nu}(k_1\cdot k_2)-k_2^\mu k_1^\nu\right]
{S}_1(m^2,Q^2,m_H^2)\epsilon_\mu(k_1)\epsilon_\nu(k_2).
\eea

\section{Transition form factor}

For a  virtual and real two-photon process, we define the
transition form factors  $F_{\rm total}$, $F_{1/2}$ and $F_{1}$ as follows:
\bea
&&S_1(m^2,Q^2,m_H^2)/\Bigl(\frac{ge^2}{(4\pi)^2}\frac{1}{m_W}\Bigr)=F_{\rm total}(Q^2,m_H^2)\nn\\
&&\qquad \quad =\sum_f N_c e_f^2 F_{1/2}(\rho_f, \tau_f)+F_1(\rho_W, \tau_W),\label{DefinitionForm}
\eea
where $e$ and $g$ are the electromagnetic and 
weak gauge couplings, respectively, and $m_W$ is
the  $W$ boson mass. $F_{1/2}$ and  $F_{1}$ are contributions from fermion loops and $W$ boson loops, respectively,  
$N_c$ is a color factor (1 for leptons and 3 for quarks), $e_f$ is the electromagnetic charge of fermion in the unit of proton charge and
\bea
\rho_f\equiv \frac{Q^2}{4m_f^2},\quad
\tau_f\equiv\frac{4m_f^2}{m_H^2}, \quad \rho_W\equiv \frac{Q^2}{4m_W^2},\quad
\tau_W\equiv\frac{4m_W^2}{m_H^2}~.
\eea
\subsection{Fermion-loop contribution}
\begin{figure}[hbt]
\begin{center}
\includegraphics[scale=0.25]{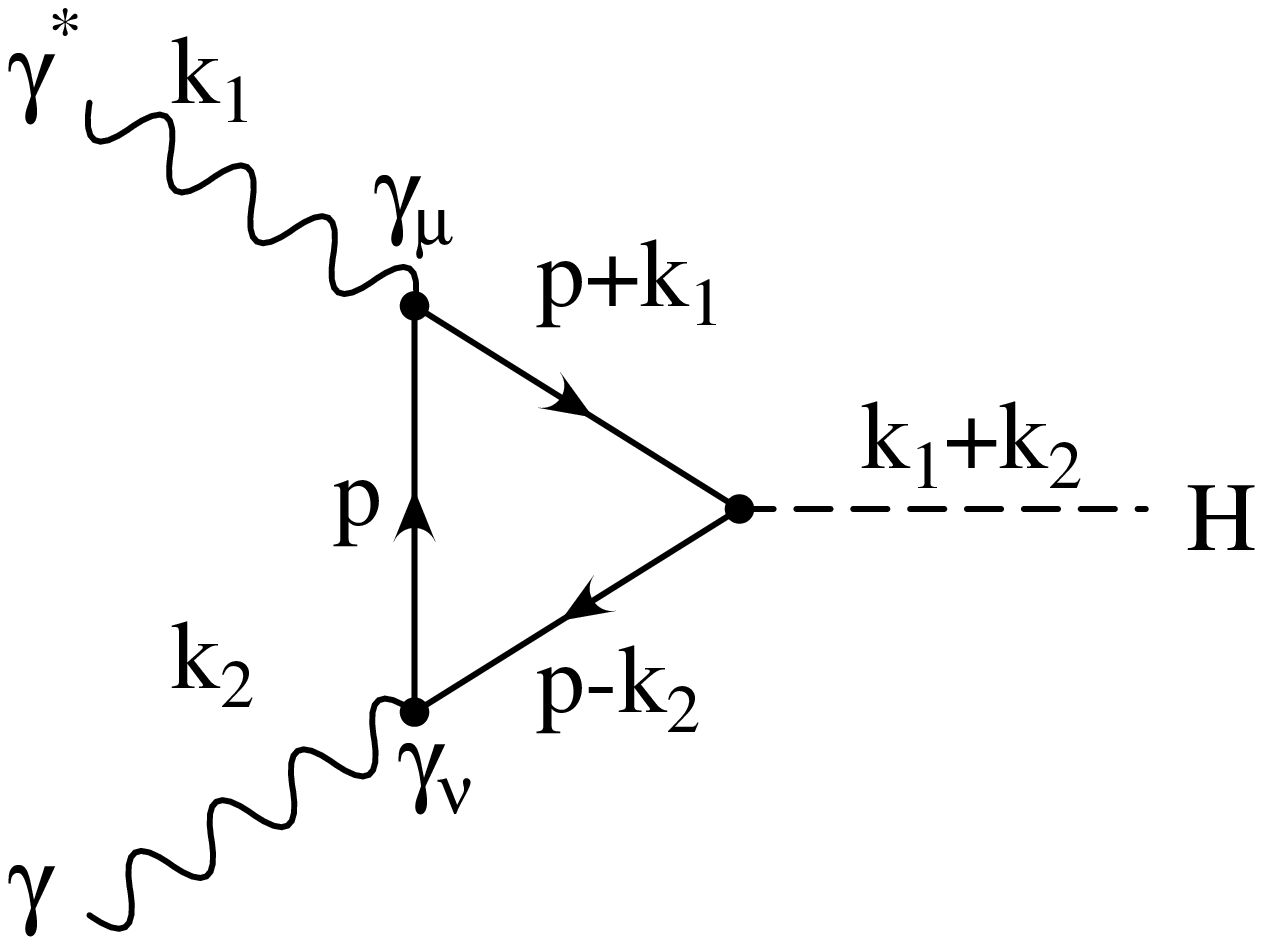}
\quad
\includegraphics[scale=0.25]{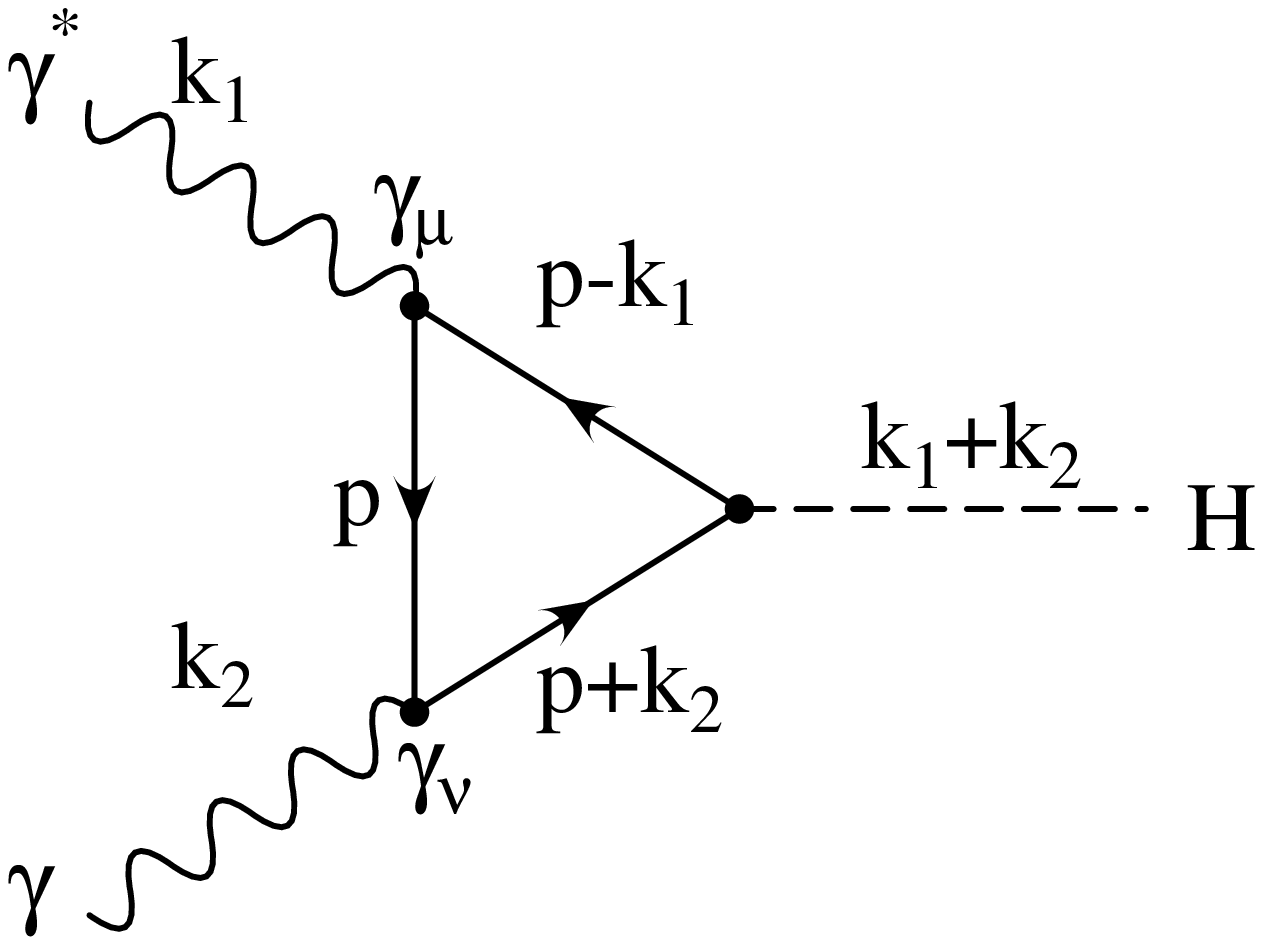}
\caption{\label{quark-loop} 
Fermion triangle-loop contribution for $\gamma^*\gamma\rightarrow H$
} 
\end{center}
\end{figure}
We calculate the
charged fermion triangle-loop diagrams shown in Fig.3 and obtain

\bea
&&\hspace{-1.2cm}F_{1/2}(\rho,\tau)=-\frac{2\tau}{1+\rho\tau}\left\{
1+\left(1-\frac{\tau}{1+\rho\tau}\right)
\left(f(\tau)+\frac{1}{4}g(\rho)\right)
\right.\nn\\
&&\hspace{-0.5cm}
\left.+\frac{\tau}{1+\rho\tau}\left(2\rho\sqrt{\tau-1}\sqrt{f(\tau)}-
\sqrt{\rho(\rho+1)}\sqrt{g(\rho)}\right)\right\}, \label{Fhalf}
\eea

where
\bea
&&\hspace{-0.5cm}f(\tau)=\left[\sin^{-1}\sqrt{\frac{1}{\tau}}\right]^2,\
\ {\rm for}\quad \tau\geq 1,\label{ftau}\\
&&\hspace{0.2cm}=-\frac{1}{4}\left[\log\frac{1+\sqrt{1-\tau}}{1-\sqrt{1-\tau}}-i\pi\right]^2
\ {\rm for}\quad \tau < 1,\\
&&\hspace{-0.5cm}g(\rho)=\left[\log
\frac{\sqrt{\rho+1}+\sqrt{\rho}}{\sqrt{\rho+1}-\sqrt{\rho}}
\right]^2. \label{grho}
\eea

Eq.(\ref{Fhalf}) shows that the fermion loop contribution $F_{1/2}$ is proportional to $\tau_f$, i.e.,  the fermion 
mass squared $m_f^2$. Thus the contributions to the transition form factor from leptons and light-flavour ($u$, $d$, $s$, $c$ and $b$) quarks are negligibly small compared to the one from top quark. 
Therefore, from now on, we consider only the top quark loop contribution for $F_{1/2}$.

\subsection{W-boson loop contribution}
\begin{figure}[hbt]
\begin{center}
\includegraphics[scale=0.25]{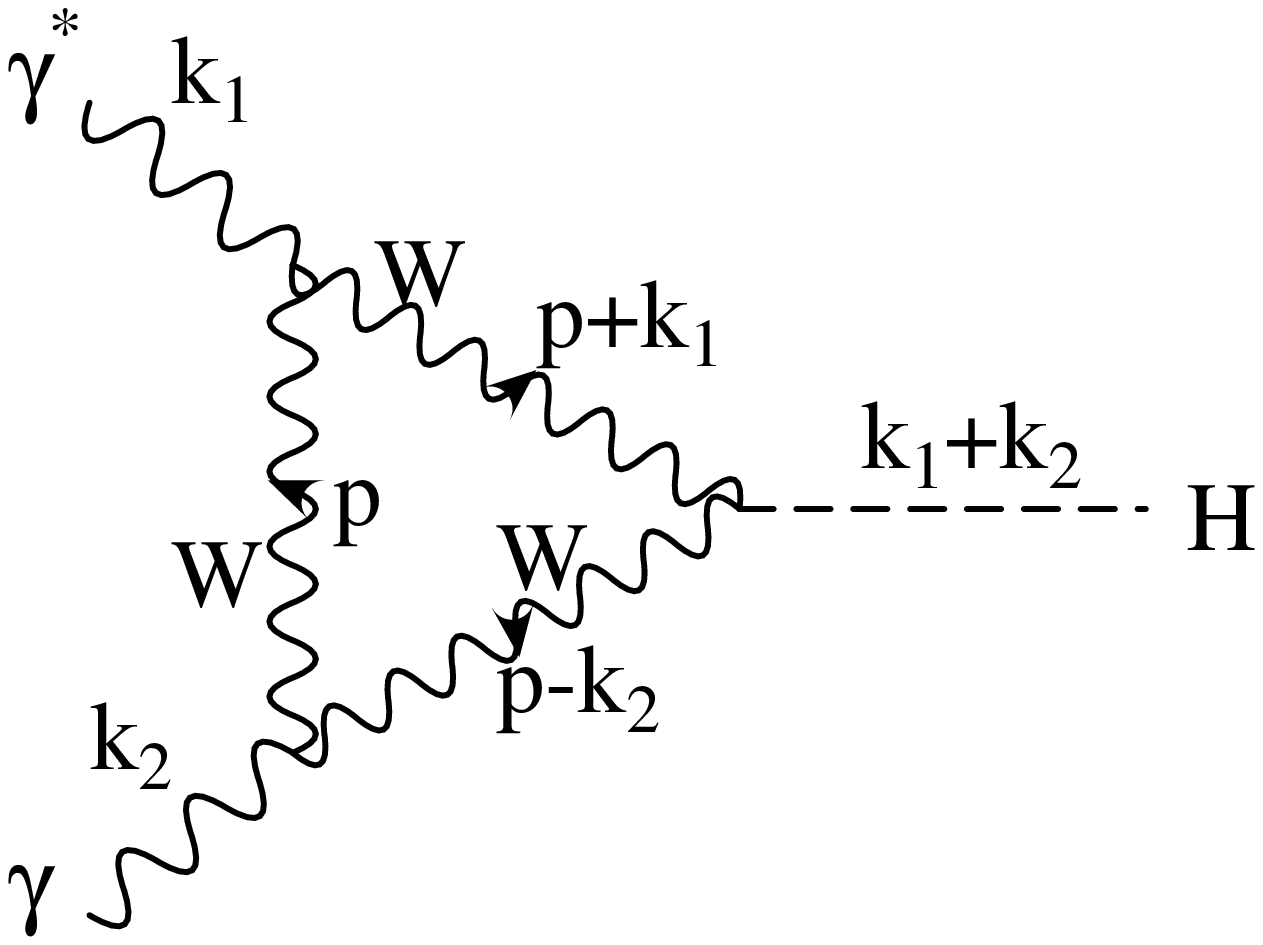}
\quad
\includegraphics[scale=0.25]{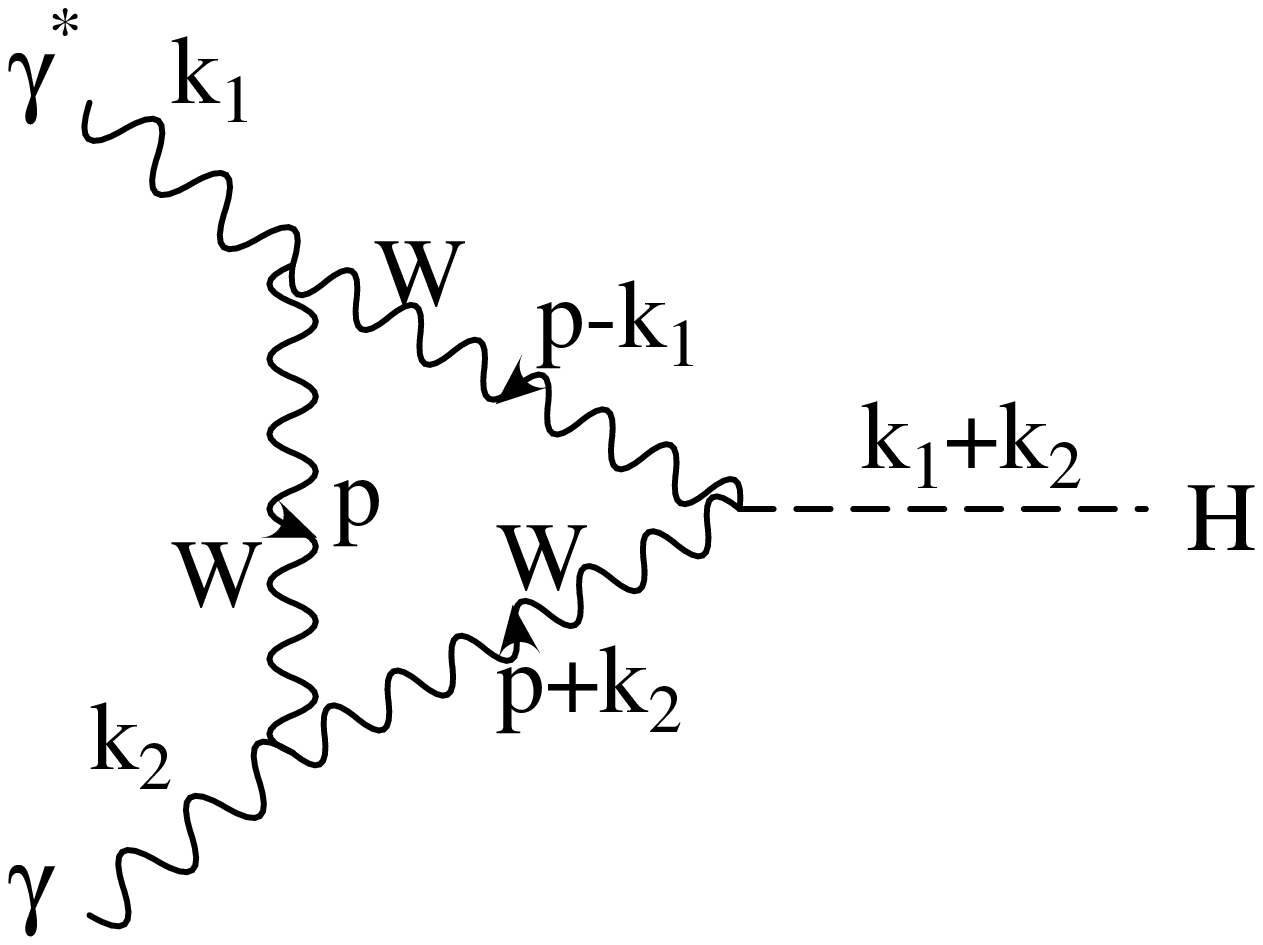}
\quad 
\includegraphics[scale=0.25]{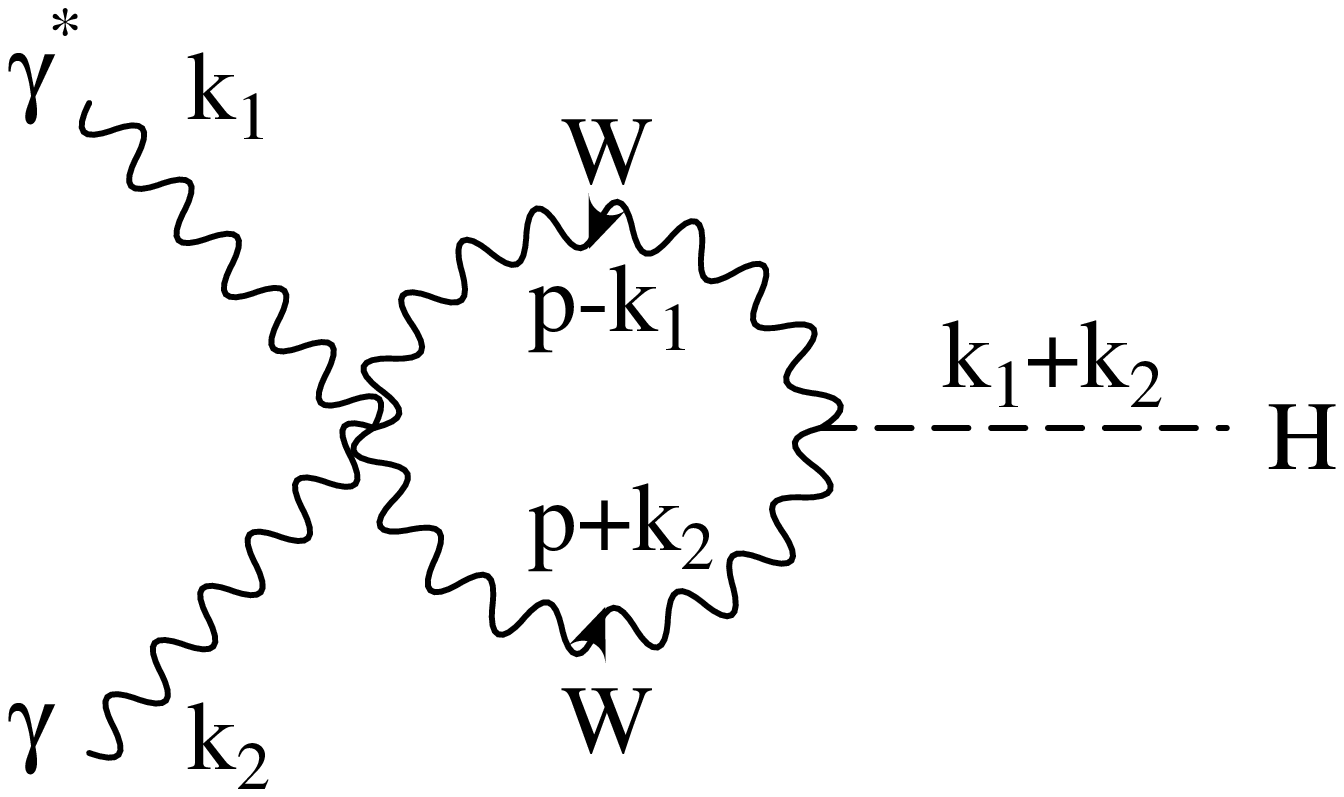}
\caption{\label{W-loop} 
W-boson loop contribution for $\gamma^*\gamma\rightarrow H$
} 
\end{center}
\end{figure}
Next we calculate the W-boson loop diagrams in unitary gauge shown 
in Fig.\ref{W-loop} and obtain
\bea
&&\hspace{-1.5cm}F_{1}(\rho,\tau)\nn\\
&&\hspace{-1.5cm}=\frac{1}{1+\rho\tau}\left\{
\frac{\tau}{1+\rho\tau}\left(
4\rho+8\rho^2\tau+6(1+\rho\tau)-3\tau\right)
\left(f(\tau)+\frac{1}{4}g(\rho)\right)\right.\nn\\
&&\hspace{-1.2cm}\left.+\left(4\rho+2(1+\rho\tau)+3\tau\right)\right.\nn\\
&&\hspace{-1.4cm}\left.\times
\left(1+\frac{2\rho\tau}{1+\rho\tau}
\sqrt{\tau-1}\sqrt{f(\tau)}-\frac{\tau}{1+\rho\tau}\sqrt{\rho(\rho+1)}
\sqrt{g(\rho)}\right)\right\},\label{Fone}
\eea
where the expressions of $f(\tau)$ and $g(\rho)$ are given in Eqs.(\ref{ftau}) and (\ref{grho}), respectively.

It is noted that $f(\tau)$ and $g(\rho)$ appear in the expressions of $F_{1/2}(\rho,\tau)$ in Eq.(\ref{Fhalf}) and $F_{1}(\rho,\tau)$ in Eq.(\ref{Fone})
 as  the following combinations, 
 \bea
\hspace{-0.5cm}f(\tau)+\frac{1}{4}g(\rho)\quad {\rm and} \quad 
2\rho\sqrt{\tau-1}\sqrt{f(\tau)}-\sqrt{\rho(\rho+1)}\sqrt{g(\rho)}~,
\eea
which arise from the 
two- and three-point scalar integrals by Passarino and 
Veltman~\cite{PV}, as will be discussed in a separate paper~\cite{WKUS}.
Similar combinations for the time-like virtual mass, which are different
from our space-like case, appear in
the Higgs decay processes $H\rightarrow \gamma^*\gamma$ and 
$H\rightarrow Z^*\gamma$ in Ref.~\cite{RomaoAndringa} 
(see also Ref.~\cite{Hunter} 
for on-shell decays, $H\rightarrow \gamma\gamma$ and $H\rightarrow Z\gamma$) .

\section{Numerical analysis}
\subsection{Transition form factor}
In the limit $Q^2\rightarrow 0$  (or $\rho\rightarrow 0$ ), $F_{1/2}(\rho_t,\tau_t)$ and 
$F_{1}(\rho_W,\tau_W)$ reduce, respectively, to
\bea
&&\hspace{-1cm}F_{1/2}(\rho_t\rightarrow 0,\tau_t)=-2\tau_t[1+(1-\tau_t)f(\tau_t)],\\
&&\hspace{-1cm}F_{1}(\rho_W\rightarrow 0,\tau_W)=2+3\tau_W+3\tau_W(2-\tau_W)f(\tau_W)~,
\eea
which coincide with the  results for the top-quark and W-boson loop contributions to $H\rightarrow 2\gamma$ decay amplitude\cite{Ellis,Ioffe,Shifman,Rizzo,Gavela, Marciano, Hunter}. 

On the other hand, for large $Q^2$, $F_{1/2}(\rho_t,\tau_t)$ and 
$F_{1}(\rho_W,\tau_W)$ show quite different behaviours:

\bea
&&\hspace{-0.8cm}
F_{1/2}(\rho_t\rightarrow \infty,\tau_t) =-\frac{1}{2\rho_t}g(\rho_t)
=-\frac{2m_t^2}{Q^2}\log^2\frac{Q^2}{m_t^2},\\
&&\hspace{-0.8cm}F_{1}(\rho_W\rightarrow \infty,\tau_W) =2g(\rho_W)
=2\log^2\frac{Q^2}{m_W^2},\eea
where we note that $g(\rho)\rightarrow \log^2(4\rho) 
\ {\rm as}\ \rho\rightarrow \infty$. Thus $F_{1/2}(\rho_t,\tau_t)$ is decreasing to zero 
while $F_{1}(\rho_W,\tau_W)$ is increasing as $Q^2$ becomes large.

\begin{figure}[hbt]
\begin{center}
\includegraphics[scale=0.75]{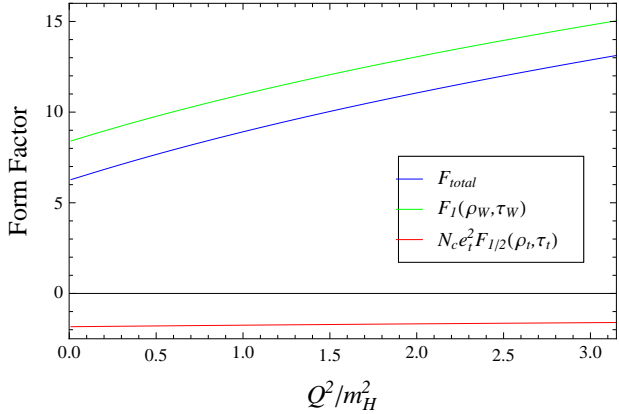}
\caption{\label{totalFF} 
Transition form factors as a function $\frac{Q^2}{m_h^2}$. Red, green and blue curves
correspond to top-quark, W-boson and total contributions, respectively. We choose mass parameters as  $m_H=126~{\rm GeV}$, $m_t=173~{\rm GeV}$ and $m_W=80~{\rm GeV}$.} 
\end{center}
\end{figure}

We plot, in Fig.\ref{totalFF},  $N_ce_t^2F_{1/2}(\rho_t,\tau_t)$, $F_{1}(\rho_W, \tau_W)$ and $F_{\rm total}$ as a function of $\frac{Q^2}{m_H^2}$. We take mass parameters as  $m_H=126~{\rm GeV}$, $m_t=173~{\rm GeV}$ and $m_W=80~{\rm GeV}$ so that we have $\tau_t=7.54$ and $\tau_W=1.61$.
We see that $W$-loop contribution $F_{1}(\rho_W, \tau_W)$ is positive,  
much larger in magnitude than top-quark loop contribution $F_{1/2}(\rho_t,\tau_t)$ and grows with $Q^2$. 
In contrast, $F_{1/2}(\rho_t,\tau_t)$ is negative and does not vary much with $Q^2$ and  stays almost constant. Thus, $F_{\rm total}$,  the sum of top-quark and $W$-boson loop contributions, grows with 
$Q^2$. Actually it grows as $\log^2\frac{Q^2}{m_W^2}$ for large $Q^2$. 

\subsection{Differential cross section}

The differential cross section for the Higgs production via $\gamma^*\gamma$ fusion in  $e\gamma \rightarrow eH$ shown in Fig.\ref{egammacollision} is given by
\bea
\hspace{-0.5cm}
\frac{d\sigma_{(\gamma^*\gamma\ {\rm fusion})}}{dQ^2}
=\frac{\alpha_{\rm em}^3}{64\pi}
\frac{g^2}{4\pi}
\frac{1}{Q^2}
\left[1+\frac{u^2}{s^2}
\right]
\frac{1}{m_W^2}|F_{\rm total}(Q^2)|^2,
\eea
where $s=(l+k_2)^2$, $u=(k_2-l')^2$ and $\alpha_{{\rm em}}=e^2/(4\pi)$.

Now a question may be posed about feasibility of extracting  
the transition form factor from  the  differential cross section.
A possible competing process for $e\gamma \rightarrow eH$ with 
single electron 
tagging is $Z\gamma$ fusion process which is obtained in Fig.\ref{egammacollision} 
by replacing the virtual photon $\gamma^*$ 
with the $Z$-boson.

\begin{figure}[hbt]
\begin{center}
\includegraphics[scale=0.75]{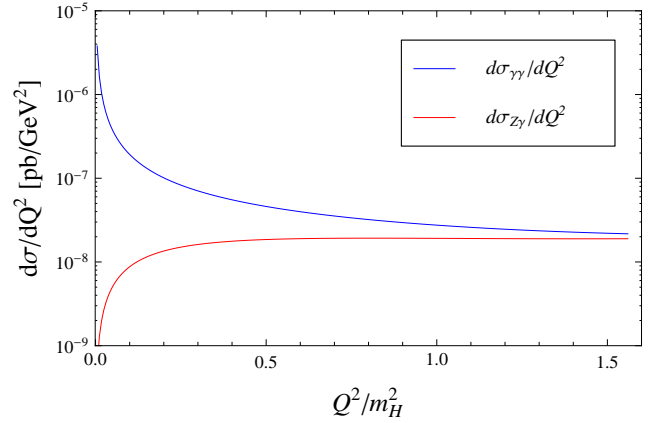}
\caption{
The differential cross section for the Higgs production as a function of
$Q^2/m_H^2$ with $m_H=126~{\rm GeV}$.
Blue and red curves
correspond to the contributions from  $\gamma^*\gamma$ fusion  and 
 $Z\gamma$ fusion, respectively. } 
\label{fig6}
\end{center}
\end{figure}

We plot,  in Fig.\ref{fig6},  the differential cross section $d\sigma/dQ^2$ for $e\gamma\rightarrow eH$ which originates from
 $\gamma^*\gamma$~($Z\gamma$) fusion
in blue~(red) line as a 
function of $Q^2/m_H^2$ for $\sqrt{s}=200$GeV. 
 Because of the mass 
squared term in the Z-boson propagator as well as  $Q^2$
which appears as an overall factor, we see that
the $Z\gamma$ fusion gives much less contribution to $d\sigma/dQ^2$ than the $\gamma^* \gamma$ fusion
in the forward region where
$Q^2/m_H^2$ is smaller than 1. 
Therefore, the transition form factor of the Higgs boson via $\gamma^* \gamma$ fusion
is measurable once a $e\gamma$ colliding machine is constructed.
In addition, regarding
the total cross section, the $\gamma^* \gamma$ fusion contribution is dominant over
 that of $Z\gamma$ fusion  for $\sqrt{s} \leq 400$ GeV, which will be 
discussed in the ref.\cite{WKUS}.

\section{Summary and Discussion}
In this paper we have studied the transition form factor of the SM
Higgs boson which arises from  top-quark loops and  $W$-boson loops. 
Its $Q^2$ dependence is summarized in Fig.\ref{totalFF}. The main contribution comes from 
$W$-boson loops and $F_{\rm total}(Q^2,m_H^2)$ grows with $Q^2$. This is a 
prediction based on the SM about the behazion form factor of the 
Higgs boson.
Any deviation of $Q^2$ dependence from the SM prediction may suggest a possible signature of the new physics
beyond  SM, such as MSSM~\cite{HaberKane} or composite models~\cite{FarhiSusskind,KaplanGeorgiDimopoulos}.

The transition form factor of the Higgs boson  may also be measured  in the
electron-positron collision, where the dominant processes for the Higgs production   are 
the Higgs-strahlung via $s$-channel $Z$-boson  and $ZZ$ or $WW$ fusion at tree
level. Consider the case in which 
one of the scattered electron (or positron) is detected (single tagging) and 
the untagged lepton is scattered into a small angle emitting an almost-real 
photon shown in Fig.\ref{eecollision}. 
Single tagging eliminates the Higgs-strahlung and  $WW$ fusion contributions, only leaving 
$ZZ$ fusion process. However, if  the kinematical region is  restricted to the  forward directions,
$ZZ$ fusion is expected to be insignificant compared to $\gamma^*\gamma$ fusion.
Then using the equivalent-photon approximation \cite{twophoton}, the 
corresponding Higgs production cross section can be written in terms of the transition form factor
given in Eq.(\ref{DefinitionForm}).

As for the future subject we should include the higher-order
effects due to QCD and electroweak interactions. 
Also it may be  
interesting to see if  the notion of transition form factor  is applicable to the 
Higgs physics at the photon collider as discussed in Ref.~\cite{MKSZ,Muhlleitner,GinzburgKrawczyk}
and the references therein.

Finally in the case of MSSM or two-Higgs
doublet model, there exist the charged Higgs bosons $H^\pm$. 
We present the  result on a  charged scalar contribution to the 
transition form factor of the Higgs boson. 
Taking  the coupling of the charged Higgs to the neutral Higgs to be
$-gm_H^{\pm 2}/m_W$ as in Ref.\cite{Hunter}, we obtain
\bea
&&\hspace{-1cm}F_0(\rho,\tau)=\frac{\tau}{1+\rho\tau}
\left[
1-\frac{\tau}{(1+\rho\tau)}\left(f(\tau)+\frac{1}{4}g(\rho)\right)\right.\nn\\&&\hspace{-0.1cm}\left.
+\frac{\tau}{1+\rho\tau}\left(2\rho\sqrt{\tau-1}\sqrt{f(\tau)}-
\sqrt{\rho(\rho+1)}\sqrt{g(\rho)}\right)\right],
\eea
where $\rho=\frac{Q^2}{4m_H^{\pm 2}}$ and $\tau=\frac{4m_H^{\pm 2}}{m_H^2}$.
In the limit $Q^2\rightarrow 0$,  we get
\bea
F_{0}(\rho\rightarrow 0,\tau)=\tau[1-\tau f(\tau)]~,
\eea
which coincides with the last equation of (2.17) of Ref.\cite{Hunter}.
Numerical analysis of $F_0(\rho,\tau)$ by varying parameters $\rho$ and $\tau$ shows 
that the scalar loop contribution is very small compared to that of  $W$ loop.


\end{document}